# Oscillating magnetic induction defined by neutron reflection

Yu.V. Nikitenko

*Frank Laboratory of Neutron Physics, JINR, 141980, Dubna, Moscow Region, Russia*

**Abstract**

The expressions are received for neutron reflection from the wave resonator placed in an oscillating magnetic field. The conditions are defined for optimal values of the neutron wave resonator and magnetic field parameters. Numerical calculations which support theory conclusions are carried out for real systems.

Corresponding author

*E-mail address:* nikiten@nf.jinr.ru (Yu.V. Nikitenko)

# 1. Introduction

In [1] is showed the spatial distribution of static and oscillating magnetic fields can be measured in the magnetic structure with use of polarized neutrons. For the spin-flip probability of the magnetic moment $\mu$, which is placed during of time $t$ in static $B_0$ and rotating with frequency $\omega$ magnetic field $B_1$ the Rabi formula has place [2, 3]

$$P = [\omega_1^2/(\omega_1^2+(\omega_0-\omega)^2))]\sin^2[\, t(\omega_1^2+(\omega_0-\omega)^2)^{1/2}/2] \tag{1}$$

where $\omega_1 = \gamma B_1$, $\gamma = \mu/s$, s is spin of particle.

It follows at $t(\omega_1^2+(\omega_0-\omega)^2)^{1/2}/2 \ll \pi/2$ the spin-flip probability does not depend of frequency $\omega$ and is

$$P = (\omega_1 t/2)^2 \tag{2}$$

Let us estimate a value of $P$ in case of neutron which crosses the magnetic layer in direction perpendicular to layer boundaries. It follows $P \approx 10^{-14}$ for layer with thickness 1Å, magnetic field amplitude in layer 1 Gs and neutron velocity perpendicular to layer boundaries 10m/s. The question is how to measure such a small value of the neutron spin-flip probability. For increasing of the neutron spin-flip reflection coefficient and increasing such way the measurement sensitivity a magnetic layer is placed in the neutron wave resonator. A neutron wave resonator is three-layered structure in which the edge layers have more high interaction potential in comparison with the middle layer. Due to the multiple neutron reflections the neutron density in a NWR increases and that reflects in increase of second emission, which appears as result of neutron absorption (capture and (or) scattering of neutrons) [4,5]. In [6, 7] is noted and in [8] is showed a sensitivity for the magnetic non-collinear resonator can be increased at certain conditions double. Analogical case exists also for magnetic system placed in an oscillating magnetic field. In given paper the question about a measurement of an oscillating magnetic induction in thin magnetic layer is considered. The conditions are defined at which the sensitivity becomes maximal.

# 2. Reflection of neutrons from resonator structure placed in oscillating magnetic field

Let us consider the external static magnetic field $H$ is parallel oriented with respect to the z-axis and parallel oriented to the internal magnetic field $B$ in the magnetic layer which placed in neutron wave resonator. Besides, in resonator and magnetic layer exists an oscillating magnetic field characterized by a frequency $2\omega$ and oriented perpendicular with respect to the z-axis by the magnetic fields $H_1$ and $B_1$, correspondently.

Our task is a determination of the frequency dependence $B_1(\omega)$ and at familiar $H_1(\omega)$ – the permeability dependence $\mu_1(\omega) = B_1/H_1$. A linear oscillating field can be represented by sum of two fields which rotate with the same frequency of $2\omega$ in opposite directions. Only the field which rotates counterclockwise around $B$ (and $H$) influences significantly a neutron polarization. As a result, the field which rotates clockwise will not be taken into account.

The Schrodinger equation in a presence of the rotating field can be reduced to a stationary problem [1], by replacing a variable magnetic field by an effective static field. As a result we have external effective field $B_{E1} = (0, 0, H-\omega)$, effective field in non-magnetic resonator layers $B_{E2} = (H_1, 0, H-\omega)$ and effective field in magnetic layer $B_{E3} = (B_1, 0, B-\omega)$.

In case of a multilayer structure (Fig. 1) we have for the reflection amplitude operator (RAO) [9]

$$R_j = r_j + t_j R_{j-1} (1 - r_j R_{j-1})^{-1} t_j \tag{3}$$

where $R_j$ is the RAO of structure with j layers, $r_j$ is the RAO of j-layer, $t_j$ is the transmission amplitude operator (TAO) of j-layer.

The RAO and TAO of j-layer can be represented with use of a vector of Pauli matrix $\sigma$ in the form [9]

$$r_j = r_j^+ + b_j \sigma r_j^-, \quad t_j = t_j^+ + b_j \sigma t_j^- \tag{4}$$

where $b_j = B_{Ej}/B_{Ej}$, $B_{Ej} = |B_{Ej}|$, $r_j^\pm = 0.5(r(B_{Ej}) \pm r(-B_{Ej}))$, $t_j^\pm = 0.5(t(B_{Ej}) \pm t(-B_{Ej}))$, $r(\pm B_{Ej}) = [r_{int}(\pm B_{Ej})(1-e^2(\pm B_{Ej})/(1-r_{int}^2(\pm B_{Ej})e^2(\pm B_{Ej}))]$, $t(\pm B_{Ej})=e(\pm B_{Ej})[(1-r_{int}^2(\pm B_{Ej}))/(1-r_{int}^2(\pm B_{Ej})e^2(\pm B_{Ej}))]$, $e(\pm B_{Ej}) = \exp(ik(\pm B_{Ej})L_j)$, $r_{int}(\pm B_{Ej})=(k-k(\pm B_{Ej}))/(k+k(\pm B_{Ej}))$, $k(\pm B_{Ej})= (k^2 - U_j - (\pm U_{Ej}))^{1/2}$, $j =1-5$, $U_j$, $U_{Ej}$ is the nuclear and effective magnetic potentials in the quadratic neutron wave vector units, correspondently.

Finally we can represent RAO of resonator structure in analogical to expressions (4) form

$$R = R^+ + R^- \sigma \tag{5}$$

Now for the spin-flip reflection amplitude for example of initial the "+" state in final the "-" state it follows

$$R_{sf} = \langle \psi^- | R | \psi^+ \rangle = R^-_x + i R^-_y \qquad (6)$$

where $\psi^+$ and $\psi^-$ are the plus and minus spin-components of neutron wave function on direction of effective magnetic field out of a resonator.

$R_{sf}$ can be represented at $r_j \approx 0$ and $t_j^- \ll t_j^+$ for $j = 2, 3, 4$ in next view

$$R_{sf} = A/[(1-B(1+D)) \times (1-B(1-D))] \qquad (7)$$

where $B = r_1 r_5 (t_2^+ t_3^+ t_4^+)^2$, $D = 2[(T_{2X}/t^+_2 + T_{3X}/t^+_3 + T_{4X}/t^+_4)^2 + (T_{2Z}/t^+_2 + T_{3Z}/t^+_3 + T_{4Z}/t^+_4)^2]^{1/2}$, $\boldsymbol{T_j} = \boldsymbol{b_j} t^-_j$

It follows for $D$ at condition $k^2 > 2\omega$

$$D = i\chi \qquad (8)$$

where $\chi = 2\{[(U_{H1}(L_2+L_4)+U_{B1}L_3)^2 + ((U_H-U_\omega)(L_2+L_4)+(U_B-U_\omega)L_3)^2]/(k^2-2U_\omega)\}^{1/2}$

Use the relation $B = |B|\exp(i\varphi)$ and supposing the $\chi < 1$ we get

$$R_{sf} = A/\{[1 - |B|\exp(i(\varphi+\chi))] \times [1 - |B|\exp(i(\varphi-\chi))]\} \qquad (9)$$

Of (9) follow the conditions for two maximal values (peaks) of $|R_{sf}(k)|$

$$\varphi(k_1) + \chi(k_1) = 2\pi n \quad \text{and} \quad \varphi(k_2) - \chi(k_2) = 2\pi n, \qquad (10)$$

where $n = 0, 1, 2, \ldots$

At enough small $\chi$ value when the $\Delta k = k_2 - k_1$ is less than resonance width we have only one $|R_{sf}(k)|$ peak. There is the equation for z- components of effective magnetic fields at which realizes the minimal value $\chi_{min} = 2[U_{H1}(L_2+L_4)+U_{B1}L_3]/[k^2-2U_\omega)]^{1/2}$, namely

$$(H - \omega)(L_2+L_4) + (B - \omega)L_3 = 0 \qquad (11)$$

It is seen the equation (11) has place at next relations

$$(B-\omega)/(\omega-H) = (L_2+L_4)/L_3 \qquad (12a)$$

$$H < \omega = (H \times (L_2+L_4) + BL_3)/(L_2+L_3+L_4) < B \qquad (12b)$$

## 3. Numerical calculations

Let us take the structure $Cu(L_1)/Al(L_2)/Co(L_3)/Al(L_4)/Cu$, which represents of itself the magnetic layer of cobalt placed inside of the neutron resonator Cu/Al/Cu. Define the magnitudes of parameters. The real parts of nuclear potential of cupper, aluminum and cobalt are 172 neV, 55 neV and 107.5 neV, correspondently. Magnetic interaction potential of neutron with the magnetic field is proportional to magnetic field induction $B$ and is $U_B(neV) = 6.03B(kGs)$. With a potential $U$ is connected a critical wave vector $k_c(Å^{-1}) = 6.96 \times 10^{-3} U^{1/2}(neV)$. Further introduce more suitable the relative system of units, in which the potentials and a neutron wave vector k will be normalized by cupper potential and cupper critical wave vector $k_{c,Cu} = 9.13 \times 10^{-3} Å^{-1}$, correspondently. In this case we have $U_{Cu} = 1$, $U_{Al} = 0.32$, $U_{Co} = 0.625$, $U_B = 0.035B(kGs)$.

In Fig 2. is shown the dependence of spin-flip neutron reflection coefficient $G_{sf}(k) = k^-|R_{sf}|^2/k^+$, where $k^-$, $k^+$ are neutron wave vectors in plus and minus states, at $U_\omega - U_H = U_B - U_\omega = 0$ and $U_\omega = 10^{-4}$. Coinciding curves 1 and 2 correspond the parameters $U_{B1} = 3 \times 10^{-3}$, $U_{H1} = 0$ and $U_{B1} = 0$, $U_{H1} = 1.15 \times 10^{-5}$ that reflects to ratio $\xi_1 = U_{B1}$(curve 1) / $U_{H1}$(curve 2) = 261. The $\xi_1$ value is close to $\eta = (L_2 + L_4)/L_3 = 300$. The curve 3 corresponds to $U_{B1} = 3 \times 10^{-3}$ and $U_{H1} = 1.15 \times 10^{-5}$ and how follows of Eq. 2 in four times exceeds the curves 1 and 2. The curves 4, 5 and 6 correspond to values $U_{B1} = 3 \times 10^{-2}$, $U_{H1} = 0$ and $U_{B1} = 0$, $U_{H1} = 1.15 \times 10^{-4}$ and $U_{B1} = 3 \times 10^{-2}$, $U_{H1} = 1.15 \times 10^{-4}$ which are in ten times bigger than ones for curves 1, 2 and 3. We see now a formation of two peaks for these curves as shown by Eq. (7).

In Fig 3. is presented the dependence of spin-flip neutron reflection coefficient $G_{sf}(k)$ at two values $L_1 = 0$ (curves 1 and 2) and $L_1 = 700 Å$ (curves 3 and 4). At $L_1 = 700 Å$ realizes the enhancement of coefficient reflection. For case $U_{B1} = 3 \times 10^{-3}$ and $U_{H1} = 1.15 \times 10^{-5}$ (curves 1 and 3) the enhancement is $\chi(700 Å) = G_{sf}(L_1 = 700Å)/ G_{sf}(L_1 = 0) \approx 2 \times 10^6$. For case $U_{B1} = 3 \times 10^{-2}$ and $U_{H1} = 1.15 \times 10^{-4}$ (curves 2 and 4) the enhancement is smaller and does not exceed $5 \times 10^4$.

In Fig. 4 is presented the $G_{sf}(k)$ dependence at different signs of $U_B - U_\omega$ and $U_H - U_\omega$. The 1 and 2 curves have one maximal value and correspond to $U_B - U_\omega = 3 \times 10^{-2}$, $U_H - U_\omega = 0$ and $U_B - U_\omega = 0$, $U_H - U_\omega = 0.9 \times 10^{-4}$. At that the $\xi_2 = U_B - U_\omega$ (curve 1)/ $U_H - U_\omega$ (curve 2) = 330 ≈ $\eta$. For curve 3 the $U_B - U_\omega = 3 \times 10^{-2} > 0$ and $U_H - U_\omega = 0.9 \times 10^{-4} > 0$ and we see increasing of distance between peaks. For curve 4 the $U_B - U_\omega = 3 \times 10^{-2} > 0$ and $U_H - U_\omega = -0.9 \times 10^{-4} < 0$ and for curve 5 the $U_B - U_\omega = -3 \times 10^{-2} < 0$ and $U_H - U_\omega = 0.9 \times 10^{-4} > 0$. We see at the different signs $U_B - U_\omega$ and $U_H - U_\omega$ the curves 4 and 5 have one maximum. Such behavior of the $G_{sf}(k)$ is connected with realization of the Eq. (11).

Estimate the minimal value $B_{1min}$, which can be measured at enhancement $2 \times 10^6$. For that define the minimal value $G_{min} = 10^{-6}$. If an oscillating field is only in the magnetic layer with thickness 1Å then have $U_{B1}= 10^{-5}$ that corresponds to $B_{1min} \approx 0.3$ Gs. Of course for that need to provide correspondent resolution of wave vector $\delta k$. In given case $2\delta k/k$ is $10^{-4}$. With increasing of the $L_1$ increases the enhancement and decreases the $\delta k/k$. So for instance at $L_1 = 900$Å the $\chi=2\times10^7$, $2\delta k/k = 10^{-5}$ and $B_{1min} \approx 0.1$ Gs. If an oscillating field $H_1$ exists in layers of resonator also then that to keep high sensitivity must be $B_1/H_1 = \mu_1 > \eta$. It is clear that to measure with high sensitivity the dependence $B_1(\omega)$ need to change the $\omega$ and the $H$ simultaneously (look the relations 13-14).

## 4. Conclusion

So it is shown with the neutron wave resonator a sensitivity of oscillating magnetic induction measurements can be increased significantly at certain conditions.

## Acknowledgment

The work is supported by RBRF grant 08-02-00467a. Author thanks to V.K. Ignatovich for fruitful discussion.

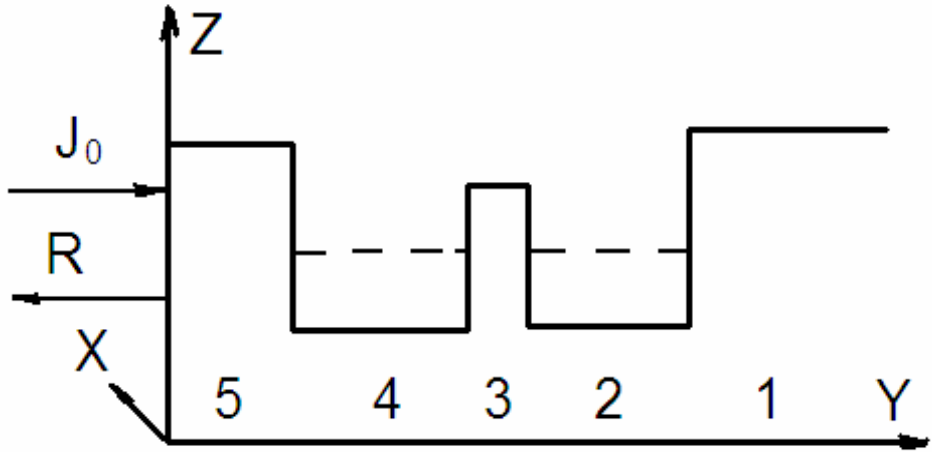

Fig. 1. Spatial dependence of the neutron nuclear interaction potential for resonator structure.

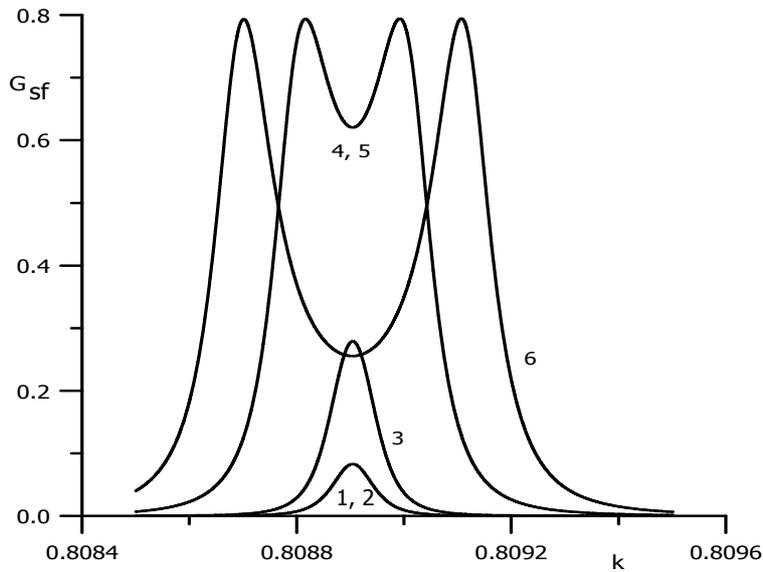

**Fig. 2.** The dependence $G_{sf}(k)$ for structure Cu/Al(150 Å)/Co(1Å)/Al(150Å)/Cu(700A) at $U_\omega - U_H = U_B - U_\omega = 0$, $U_\omega = 10^{-4}$ and $U_{B1}$, $U_{H1}$ values: $U_{B1} = 3\times10^{-3}$, $U_{H1} = 0$ (curve 1); $U_{B1} = 0$, $U_{H1} = 1.15\times10^{-5}$ (curve 2);: $U_{B1} = 3\times10^{-3}$, $U_{H1} = 1.15\times10^{-5}$ (curve 3); $U_{B1} = 3\times10^{-2}$, $U_{H1} = 0$ (curve 4); $U_{B1} = 0$, $U_{H1} = 1.15\times10^{-4}$ (curve 5); $U_{B1} = 3\times10^{-2}$, $U_{H1} = 1.15\times10^{-4}$ (curve 6).

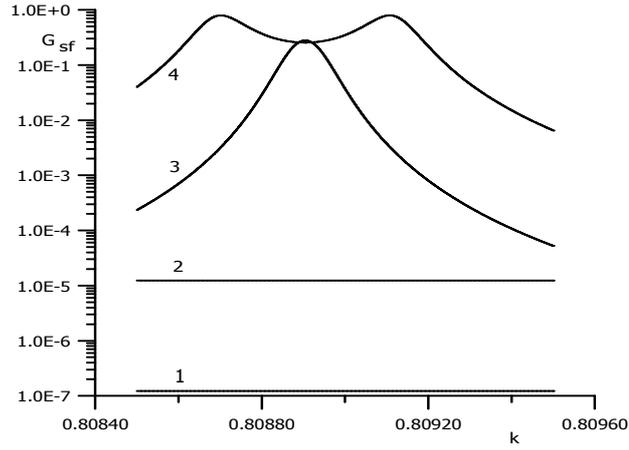

**Fig 3.** The dependence $G_{sf}(k)$ for structure Cu/Al(150 Å)/Co(1Å )/Al(150Å)/Cu($L_1$) at $U_\omega - U_H = U_B - U_\omega = 0$ and $U_\omega = 10^{-4}$. $L_1 = 0$ (curves 1 and 2) and $L_1 = 700$Å (curves 3 and 4); $U_{B1} = 3\times10^{-3}$ and $U_{H1} = 1.15\times10^{-5}$ (curves 1 and 3); $U_{B1} = 3\times10^{-2}$ and $U_{H1} = 1.15\times10^{-4}$ (curves 2 and 4).

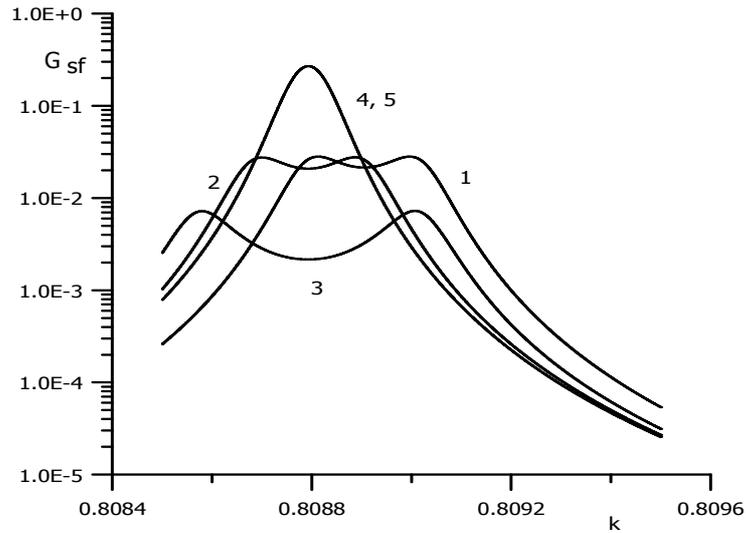

**Fig 4.** The dependence $G_{sf}(k)$ for structure Cu/Al(150 Å)/Co(1Å )/Al(150Å)/Cu(700A) at $U_{B1} = 3\times10^{-3}$, $U_{H1} = 1.15\times10^{-5}$, $U_\omega = 3\times10^{-4}$ and $U_B - U_\omega$ and $U_H - U_\omega$ values: $U_B - U_\omega = 3\times10^{-2}$ and $U_H - U_\omega = 0$ (curve 1); $U_B - U_\omega = 0$ and $U_H - U_\omega = 0.9\times10^{-4}$ (curve 2 ); $U_B - U_\omega = 3\times10^{-2}$ and $U_H - U_\omega = 0.9\times10^{-4}$ (curve 3); $U_B - U_\omega = 3\times10^{-2}$ and $U_H - U_\omega = -0.9\times10^{-4}$ (curve 4); $U_B - U_\omega = -3\times10^{-2}$ and $U_H - U_\omega = 0.9\times10^{-4}$ (curve 5) .

**Figure captions**

**Fig. 1.** Spatial dependence of the neutron nuclear interaction potential for resonator structure.

**Fig. 2.** The dependence $G_{sf}(k)$ for structure Cu/Al(150 Å)/Co(1Å)/Al(150Å)/Cu(700A) at $U_\omega - U_H = U_B - U_\omega = 0$, $U_\omega = 10^{-4}$ and $U_{B1}$, $U_{H1}$ values: $U_{B1} = 3\times10^{-3}$, $U_{H1} = 0$ (curve 1); $U_{B1} = 0$, $U_{H1} = 1.15\times10^{-5}$ (curve 2);: $U_{B1} = 3\times10^{-3}$, $U_{H1} = 1.15\times10^{-5}$ (curve 3); $U_{B1} = 3\times10^{-2}$, $U_{H1} = 0$ (curve 4); $U_{B1} = 0$, $U_{H1} = 1.15\times10^{-4}$ (curve 5); $U_{B1} = 3\times10^{-2}$, $U_{H1} = 1.15\times10^{-4}$ (curve 6).

**Fig 3.** The dependence $G_{sf}(k)$ for structure Cu/Al(150 Å)/Co(1Å)/Al(150Å)/Cu($L_1$) at $U_\omega - U_H = U_B - U_\omega = 0$ and $U_\omega = 10^{-4}$. $L_1 = 0$ (curves 1 and 2) and $L_1 = 700$Å (curves 3 and 4); $U_{B1} = 3\times10^{-3}$ and $U_{H1} = 1.15\times10^{-5}$ (curves 1 and 3); $U_{B1} = 3\times10^{-2}$ and $U_{H1} = 1.15\times10^{-4}$ (curves 2 and 4).

**Fig 4.** The dependence $G_{sf}(k)$ for structure Cu/Al(150 Å)/Co(1Å)/Al(150Å)/Cu(700A) at $U_{B1} = 3\times10^{-3}$, $U_{H1} = 1.15\times10^{-5}$, $U_\omega = 3\times10^{-4}$ and $U_B - U_\omega$ and $U_H - U_\omega$ values: $U_B - U_\omega = 3\times10^{-2}$ and $U_H - U_\omega = 0$ (curve 1); $U_B - U_\omega = 0$ and $U_H - U_\omega = 0.9\times10^{-4}$ (curve 2); $U_B - U_\omega = 3\times10^{-2}$ and $U_H - U_\omega = 0.9\times10^{-4}$ (curve 3); $U_B - U_\omega = 3\times10^{-2}$ and $U_H - U_\omega = -0.9\times10^{-4}$ (curve 4); $U_B - U_\omega = -3\times10^{-2}$ and $U_H - U_\omega = 0.9\times10^{-4}$ (curve 5).